\documentclass[11pt,a4paper]{article}
\usepackage[utf8]{inputenc}
\usepackage[english]{babel}
\usepackage{amsmath}
\usepackage{amsfonts}
\usepackage{amssymb}
\usepackage{amsthm}
\usepackage[margin=0.8in]{geometry}

\begin{document}
\newcommand{\Norm}[1]{\vert\vert#1\vert\vert} 
\newcommand{\Abs}[1]{\vert#1\vert} 
\newcommand{\Dotp}[2]{\langle#1,#2\rangle} 
\newcommand{\RR}{\mathbb{R}} 
\newcommand{\FT}[1]{\int_{\RR^n} #1 e^{-i\Dotp{k}{x}} dm(x)} 
\newcommand{\IFT}[1]{\int_{\RR^n}\hat#1 e^{i\Dotp{k}{x}} dm(k)} 

\newtheorem{theorem}{Theorem}
\newtheorem{proposition}{Proposition}
\newtheorem{corollary}{Corollary}
\newcommand{\LL}{\cal L} 

\title{On the definiteness of associated energy-momentum tensors to a class
       of general variation problems}
\author{Kurt Pagani\\{\tt pagani@scios.ch}}
\date{\today}
\maketitle
\abstract{We show the point-wise definiteness and some other properties of the energy-momentum tensor for a certain class of Euler-Lagrange equations under quite general conditions.}

\section{Introduction and Results}
Let ${\LL}:{\RR}^{n}\times\RR\rightarrow\RR$,\,$(\xi,\eta)\mapsto {\LL}(\xi,\eta)$ be a smooth function and $\Omega$ an open subset of
$\RR^n$, then a critical point $u$ of the functional
\begin{equation} \label{FUNC}
  S(\phi)=\int_{\Omega} {\LL}(\nabla\phi(x),\phi(x))\,dx 
\end{equation} 
satisfies the Euler-Lagrange equation (at least in a weak sense)
\begin{equation} \label{ELEQ}
 {\tt div}\, {\LL}_{\xi}(\nabla u(x),u(x))= {\LL}_{\eta}(\nabla u(x), u(x))
\end{equation}
assuming $u$ vanishes on $\partial\Omega$. Associated to $(\ref{ELEQ})$ is the tensor
\begin{equation} \label{TEN}
  T = \nabla u \,\otimes\ {\LL}_{\xi}(\nabla u,u) -
      {\LL}(\nabla u,u)\, {\tt Id},   
\end{equation} 
or written in components:
\begin{equation}\label{TENC}
  T_{ij}(x)=u_i(x) {\LL}_{\xi_j}(\nabla u(x),u(x)) - \delta_{ij}  {\LL}(\nabla u(x),u(x)),
\end{equation}
for $x\in\Omega,\,  1\leq i,j \leq n$. It is easily checked, using $(\ref{ELEQ})$, that
\begin{equation}\label{DIVT0}
  {\tt Div}(T)=0 \hspace{10pt} {\tt in} \, \Omega, 
\end{equation}
that is $\sum_{j=1}^n T_{ij,j}=0$ for all $i=1\ldots n$. The tensor $T$ is usually not 
symmetric but, wherever point-wise defined, it holds:
\begin{equation}
 T \nabla u = (\Dotp{{\LL_{\xi}}}{\nabla u}-{\LL}) \nabla u,
\end{equation}
that is $\nabla u(x)$ is an eigenvector to the eigenvalue 
\begin{equation}\label{EIGEN1}
\lambda_1(x)= \Dotp{{\LL_{\xi}}}{\nabla u}-{\LL}.
\end{equation}
Moreover, any vector $X$ belonging to the orthogonal complement of ${\LL_{\xi}}$ at $x$ gives
\begin{equation}
 T X = - {\LL}(\nabla u,u) X
\end{equation}
so that we have $(n-1)$ other eigenvalues 
$\lambda_2(x)=\ldots=\lambda_n(x)=-{\LL}(\nabla u(x),u(x))$. Obviously, it would be preferable
to have ${\LL_{\xi}} \parallel \nabla u$, so that $T$ were symmetric and the spectrum well
defined (none of the $X$ could be parallel to $\nabla u$). 

When we define $n$ differential $(n-1)$-forms $\omega_i$ on $\Omega$ as
\begin{equation}
   \omega_i(x) = T_{ij}(x) \star{dx_j} = \sum_{j=1}^n (-1)^{j-1} T_{ij}(x) \, 
    dx_1\wedge\ldots \wedge\widehat{dx_j}\wedge \ldots \wedge dx_n,
\end{equation}   
then $(\ref{DIVT0})$ implies
\begin{equation}
  d\omega_i=0, \hspace{20pt} i=1\ldots n,
\end{equation}
where $\star$ is the Hodge operator, with sign convention so that 
$\omega\wedge\star\nu =\Dotp{\omega}{\nu}\eta$ holds.
Thus, if $\Omega$ is such that the Poincare lemma is applicable, then there exist 
$n$ one forms $\star d\sigma_i$ on $\Omega$ satisfying
\begin{equation}
    \star d\sigma_i \wedge \ldots \wedge \star d\sigma_n = {\tt det}(T) 
    \,dx_1\wedge\ldots\wedge dx_n,
\end{equation}
leading to the question: under what conditions is $T$ definite on $\Omega$,
or non-degenerate at least? In either case the $1$-forms are linearly
independent and may be used as coordinate transforms. 
We got the following results.

\begin{theorem}
Let $\Omega$ be a bounded domain in $\RR^n$ with a $C^{2+\alpha}$-boundary, whose
mean curvature $H(y)$ is non-negative at every $y\in\partial\Omega.$ Suppose 
\[ {\LL}(\xi,\eta)=F(|\xi|,\eta), \]
where $F\in C^2(\RR_{+}\times\RR)$ is strictly convex in the first variable. Then for any 
classical solution $u\in C^3(\Omega)\cap C^{2}(\overline{\Omega})$ of $(\ref{ELEQ})$ with
boundary condition $u|_{\partial\Omega}=0$, the
following statements hold: 
  \begin{itemize}
    \item[1.] $T$ as defined in $(\ref{TEN})$ is symmetric.
    \item[2.] If $F(p,q)>0$ on $\{p\geq 0\}\times \RR$, then $T$ is negative definite on
          $\overline{\Omega}$, i.e. $\Dotp{\xi}{T(x)\xi}\leq -C |\xi|^2$, for some
          positive constant $C$ and for all $\xi\in\RR^n, x\in\ \overline{\Omega}$.\\
          If, additionally, $F$ is non-decreasing in the second variable and 
          $m\leq u(x) \leq M$ on $\overline{\Omega}$, then $C=-F(0,m)$. 
    \item[3.] If $F(p,q)<0$ and $p F_p(p,q)-F(p,q)>0$ on $\{p\geq 0\}\times \RR$, then
          $T$ is positive definite on $\Omega$. 
  \end{itemize}
Moreover, for both cases above:

  \begin{itemize}
    \item[4.] ${\tt det}(T)= (|\nabla u|F_p-F) F^{n-1}\neq 0,\, \forall x\in\Omega$
    \item[5.] ${\tt Tr}(T)=|\nabla u|F_p-n F \lessgtr 0$. 
    \item[6] $|\nabla u| F_p(|\nabla u|,u)-F(|\nabla u|,u) \leq -F(0,\zeta), $
       \hspace{2pt} $\forall x\in\Omega$ and some $\zeta\in \{u(x):\nabla u=0\}$
  \end{itemize}
  
\end{theorem}

Only the most general cases are listed. It will be obvious in the proof that there are a lot
of other possibilities. Furthermore, there is much room for improvement as soon as less
generality is required. For instance, if $F$ is a sum or product of two terms, the
conditions can be considerably relaxed. Moreover, some requirements need to be 
valid only on the ranges of the solutions, however, as those are not necessarily known a priori 
this fact was not used.  
The theorem is more or less a corollary of the following proposition.

\begin{proposition}
Let $\Omega$ be a bounded domain in $\RR^n$ with smooth boundary $\partial\Omega\in C^{2+\alpha}$. Furthermore, let
\[ {F}:{\RR\cap \{p \geq 0\} }\times\RR\rightarrow\RR ,\, (p,q)\mapsto {F}(p,q) \] 
be a $C^2$ function which is is strictly convex in the first variable. 
Setting
\[ {\LL(\xi,\eta)}=F(|\xi|,\eta), \]
then for any classical solution $u\in C^3(\Omega)\cap C^{2}(\overline{\Omega})$  
of $(\ref{ELEQ})$, vanishing on $\partial\Omega$, the eigenvalue $(\ref{EIGEN1})$ of $T$ is
given by
\begin{equation}
 \lambda_1(x) = |\nabla u| F_p(|\nabla u|,u)-F(|\nabla u|,u)
\end{equation}
and assumes its maximum either on the set $C_u=\{x:\nabla u(x)=0 \}$ or on the boundary
$\partial \Omega$. Thus
\[
   \sup_{\overline{\Omega}}{\lambda_1(x)}={\tt max} \{ -\inf_{C_u} F(0,u(x)),\,
   \sup_{\partial\Omega} ( |\tfrac{\partial u}{\partial n}|\, 
      F_p(|\tfrac{\partial u}{\partial n}|,0)-F(|\tfrac{\partial u}{\partial n}|,0)   \}
\]
If the mean curvature $H$ of $\partial\Omega$ is non-negative at every point, then
\[
   \sup_{\overline{\Omega}}{\lambda_1(x)}= -\inf_{C_u} F(0,u(x)).
\]

\end{proposition}

The reader acquainted with the so called $P$-functions introduced by Payne and
Phillipin \cite{PP2}, will easily recognize that $\lambda_1$ provides a simple recipe to get such
a function without guessing. That is: build the associated tensor and apply it to the 
gradient of the solution. However, to show the non-degeneracy or definiteness of $T$, 
additional restrictions to the function $F$ had to be imposed in order to guarantee that no
eigenvalue is zero or that all eigenvalues have the same sign. 

For later reference we state the following simple identities, valid for any 
smooth critical point of $S(\phi)$:
Let $X(x)=x-x_0$ for some $x_0\in\Omega$, then
\[ 
      {\tt div}(T X)= \Dotp{{\tt Div}(T)}{X}+{\tt Tr}(T D_x X)={\tt Tr}(T)
      =\Dotp{\nabla\phi}{{\LL}_{\xi}}- n\, {\LL},
\]
and integrating over $\Omega$ yields:
\begin{equation}\label{EQREF1}
   \int_{\Omega} \left( \Dotp{\nabla\phi}{{\LL}_{\xi}}- n {\LL} \right) dx=
   \int_{\partial\Omega} \Dotp{y-x_0}{T(y)\nu(y)} dH(y)^{n-1},
\end{equation}
and using $(\ref{ELEQ})$, it follows that  
$\Dotp{\nabla\phi}{{\LL}_{\xi}}=
{\tt div}(\phi\, {\LL}_{\xi})-\phi \,{\tt div}({\LL}_{\xi})= \phi\,{\LL}_\eta,$
therefore
\begin{equation}\label{EQREF2}
  \int_{\Omega} \left( \phi\,{\LL}_\eta - n\,{\LL} \right) dx= \sum_{i,j}
   \int_{\partial\Omega}  \left( T_{ij}(y) X_i(y)  - 
    \phi(y)\, {\LL}_{\xi_j} \right) \nu(y)_j  \, dH(y)^{n-1}.
\end{equation}

Now let us look at a simple example. Setting
\[
      {\LL}(\xi,\eta)=\tfrac{1}{2}|\xi|^2+ \Phi(\eta),\,\Phi>0
\]
then the function $F(p,q)=\tfrac{1}{2}p^2+\Phi(q)$ satisfies all the conditions if 
$\phi$ is $C^2$. The Euler-Lagrange equation reads as
\[
     \Delta u(x) = \Phi'(u(x)),\ u|_{\partial\Omega=0}
\]
and $T$ has the form:
\[
   T_{ij}= u_i(x) u_j(x)- \delta_{ij} \left( \tfrac{1}{2}|\nabla u(x)|^2+\Phi(u(x))  \right).
\]
If $u$ is as smooth as required, then the value of $\Dotp{X}{T\nu}$ on $\partial\Omega$ 
is given by:
\[
    \Dotp{X}{T\nu}=\partial_X u\, \partial_\nu u - 
    \tfrac{1}{2}\Dotp{X}{\nu} \left( |\partial_\nu u|^2 +
    \Phi(0)  \right) = \tfrac{1}{2}\Dotp{X}{\nu}
    \left( |\partial_\nu u|^2-\Phi(0) \right).
\] 
Inserting into $(\ref{EQREF1})$ yields
\[
     \int_{\Omega} \left( \tfrac{2-n}{2} |\nabla u|^2 - n\,\Phi(u) \right) dx =
       \int_{\partial\Omega}  \tfrac{1}{2}\Dotp{X}{\nu}
        \left( |\partial_\nu u|^2-\Phi(0) \right) dH^{n-1}.
\]
On the other hand, the Euler-Lagrange equation gives
\[
     \int_{\Omega} |\nabla u|^2 \,dx =  \int_{\Omega} u\, \Phi'(u) \,dx   ,
\]
so that we get finally 
\[
     \int_{\Omega} \left( \tfrac{2-n}{2} u\,\Phi'(u) - n\,\Phi(u) \right) dx +
      \tfrac{n}{2} \Phi(0) |\Omega| =
       \tfrac{1}{2}\int_{\partial\Omega}  \Dotp{X}{\nu}
         |\partial_\nu u|^2 dH^{n-1}.
\]

The expression above is the well known Rellich identity \cite{Rellich}, also known as 
Pohozaev identity, so that $(\ref{EQREF1}),(\ref{EQREF2})$ are a generalization
of this to any Euler-Lagrange equation. When one recollects that the energy-momentum
tensor $T$ is a consequence of the variation of $\Omega$ by the diffeomorphisms
\[
   x \mapsto x+\epsilon\,S(x),\,\, S\in C_0^\infty(\Omega,\RR^n)
\] 
then (well known as Noethers theorem) the equation $(\ref{DIVT0})$ is the expression
for a conservation law which imposes some restrictions on ${\LL}$ and/or $\Omega$ in order to
have solutions at all. Indeed, if, for instance, $\Omega$ is star-shaped, then the
quantity $\Dotp{X}{\nu}$ is non-negative, so that it is easy to find $\Phi$ for which there
is no solution.
Now, by Proposition 1,
\[
    \lambda_1(x) = \tfrac{1}{2} |\nabla u(x)|^2 - \Phi(u(x))
\]
assumes its maximum either on the boundary of $\Omega$ or where $u$ has a critical point. 
If, for example, $\Omega$ is convex, then the mean curvature is certainly non-negative, so that
the maximum is attained at a point where $\nabla u=0$. With the additional
assumption $\Phi' \geq 0$, it follows by the maximum principle that 
\[
             m \leq u(x) < 0 \,\,\, in \,\,\Omega,
\] 
thus 
\[
    \lambda_1(x) = \tfrac{1}{2} |\nabla u(x)|^2 - \Phi(u(x)) \leq -\Phi(m)
\]
yielding the gradient bound
\[
          \tfrac{1}{2} |\nabla u(x)|^2  \leq  \Phi(u(x)) - \Phi(m), \forall x\in\Omega.         
\]

The example above is almost typical for the general case. For further examples and 
applications we refer to \cite{Sperb} and \cite{PP2}.

\section{Proofs}
To prove Proposition 1, we will use the maximum principle of Payne-Philippin 
\cite{PP} since a
direct proof - although feasible - does not reveal any new facts. The principle states that,
if $u$ is a classical solution of

\begin{equation}\label{PPEQ}
  {\tt div} (g(|\nabla u|^2,u) \nabla u)+h(|\nabla u|^2,u)=0 \hspace{8pt} in\,\,\Omega\subset 
  {\RR^n}
  \end{equation}  
with $u|_{\partial\Omega}=0$, $g\in C^1,\,h\in C^0$, and 
\begin{equation}\label{PPC1}
   g(p^2,q)+ 2p^2  \frac{\partial g}{\partial p^2}(p^2,q) > 0
\end{equation} 
then any solution $\Phi$ of

\begin{equation}\label{PPC2}
  2\left( h(p^2,q)+p^2 \frac{\partial g}{\partial q}(p^2,q) \right)\, 
    \frac{\partial \Phi}{\partial p^2}=
   \left( g(p^2,q)+ 2p^2  \frac{\partial g}{\partial p^2}(p^2,q) \right) \,  
     \frac{\partial \Phi}{\partial q}   
\end{equation}
satisfying
\begin{equation}\label{PPC3}
   \frac{\partial \Phi}{\partial p^2} > 0
\end{equation}
assumes its maximum value either on $\partial\Omega$ or at a critical point of $u$, whereby
$$\Phi(p^2,q)|_u=\Phi(|\nabla u(x)|^2,u(x)).$$ Moreover, if the mean curvature of 
$\partial\Omega$ is non-negative, then $\Phi$ cannot assume a maximum value 
on $\partial\Omega$.

So, setting
\[
   g(p^2,q)=\frac{F_p(p,q)}{p},\hspace{10pt} h(p^2,q)=-F_q(p,q)
\]
and
\begin{equation}\label{PPSOL}
   \Phi(p^2,q) = p\,F_p(p,q)-F(p,q),
\end{equation}
we only have to show that $(\ref{PPC1}),(\ref{PPC2})$ and $(\ref{PPC3})$ are satisfied, then
$(\ref{PPSOL})$ is a solution which is just $\lambda_1$ when evaluated at $u$. 
Beginning with $(\ref{PPC1})$, we get
\[
     \frac{F_p(p,q)}{p}+ 2 p^2   
       \frac{\partial}{\partial p} \left(  \frac{F_p(p,q)}{p} \right) 
       \frac{\partial p}{\partial p^2} = F_{p p},
\]
which is positive by the strict convexity of $F$ and the regularity assumption. So this point
- none but the ellipticity condition for $(\ref{PPEQ})$ - is satisfied. Next, $(\ref{PPC2})$
reads
\[
 2 \left( -F_q+p F_{p q}  \right) \Phi_{p^2}=F_{pp} \Phi_q, 
\]
and inserting $(\ref{PPSOL})$ yields
\[
   2 \left( -F_q+p F_{p q}  \right) \left( F_p+p\,F_{pp} -F_p \right)\frac{1}{2p}=
   F_{pp} \left( p F_{pq} - F_q \right),
\] where all terms cancel. It remains $(\ref{PPC3})$, i.e. 
\[
    \Phi_{p^2}=\frac{1}{2p} \left(  F_p+pF_{pp}- F_p  \right)=F_{pp} > 0
\]
as required. So Proposition 1 is proven when setting
\[
    \lambda_1(x)=\Phi(p^2,q)|_u = |\nabla u(x)|\, F_p(|\nabla u|,u) - F(|\nabla u|,u).
\]
Indeed, $(\ref{TENC})$ has the form
\[
    T_{ij}= \frac{F_p(|\nabla u|,u)}{|\nabla u|} u_i \, u_j - \delta_{ij} F(|\nabla u|,u),
\]
thus
\[
    T_{ij} u_j = \left(  |\nabla u(x)|\, F_p(|\nabla u|,u) - F(|\nabla u|,u) \right)\, u_i(x)
    = \lambda_1(x) \,u_i(x).
\]

To prove Theorem 1, recall that the eigenvalues of $T$ are given by
\[
     \lambda_1(x)=|\nabla u|\,F_p(|\nabla u|,u)-F(|\nabla u|,u),\, \lambda_j=-F(|\nabla u|,u),
     \,\, j=2,\ldots,n,
\]
and that $T$ is symmetric (as seen above). The supposition on $\partial\Omega$ implies that
Proposition 1 applies and that $\lambda_1$ assumes its maximum value at a critical point of
$u$. Therefore all statements are straightforward consequences of those facts. For the
definiteness of $T$ we must merely assure that all eigenvalues have the same sign. 
The uniformity of the definiteness follows by the regularity of the solution up to the 
boundary and the compactness of the closure of $\Omega$. As already remarked, the more
one specializes the Lagrangian $F$ the more accurate information one gets. 
To conclude the proof, we have to show that the regularity requirements to the solution
$u$ are sufficient. 

For this purpose we have to look at the method of proof for $P$-functions in general. Recall
$(\ref{EIGEN1})$:
\[
      \lambda_1(x)= \Dotp{{\LL_{\xi}}}{\nabla u}-{\LL}.
\] 
The general method is to show that $\lambda_1$ satisfies a differential inequality 
(usually of second order) such that the classical maximum principle applies
or that
\[
   \nabla \lambda_1 = ({\LL}_{\xi \xi}\, D^2 u)\nabla u+ 
   \Dotp{{\LL}_{\xi \eta}}{\nabla u} \nabla u
   - {\LL}_{\eta} \nabla u = M \nabla u
\]
can vanish only at a point where $\nabla u=0$ in case $\lambda_1$ assumes a maximum there.
In either case we need $C^2$ regularity of $u$ up to the boundary. To decide whether there
is an interior maximum or not and also for a second order inequality of $\lambda_1$ we have
to know the second derivatives $D^2 \lambda_1$ which involve the third derivatives of $u$.
Therefore the minimal regularity conditions in Theorem 1 are sufficient. In certain
cases it is surely possible to approximate less regular solutions by the method of
continuation (\cite{GT}) but a general theory in this generality seems to be difficult. Usually,
higher regularity is easy to achieve when one has shown existence and classical
regularity (i.e $C^{2+\alpha}$) of a solution. For this and other related topics we
refer to Gilbarg-Trudinger \cite{GT}, Chapters 10,15.
To conclude, we remark that in the case of convex solutions ($D^2u>0$) it is possible
to get a much simpler theory even for non-elliptic ${\LL}'s$.

\end{document}